\documentclass[aps,showpacs,pra,superscriptaddress,]
{revtex4-2}
\usepackage{amsmath}
\usepackage{amsfonts}
\usepackage{amssymb}
\usepackage{bm}
\usepackage{graphicx}

\begin{document}

\title{Stopping of generalized solitary and periodic waves in optical
waveguide with varying and constant parameters}
\author{Vladimir I. Kruglov}
\affiliation{Centre for Engineering Quantum Systems, School of Mathematics and Physics,
The University of Queensland, Brisbane, Queensland 4072, Australia}
\author{ Houria Triki}
\affiliation{Radiation Physics Laboratory, Department of Physics, Faculty of Sciences,
Badji Mokhtar University, P. O. Box 12, 23000 Annaba, Algeria}

\begin{abstract}
We demonstrate the possibility of stopping the soliton pulses in optical waveguide with varying and constant parameters exhibiting a Kerr nonlinear response. By using the similarity transformation, we have found the constraint condition for varying waveguide parameters and derive the exact analytical
solutions for self-similar bright, kink, dark and rectangular solitary and periodic waves for nonlinear Schr\"{o}dinger equation with  variable coefficients. All these generalized wave solutions depend on five arbitrary parameters and two free integration constants.
It is found that the velocity of solitons is related to a free parameter $q$, which play an important role in the dynamic behavior of soliton's evolution.
The precise expression of soliton's velocity shows that the
solitons can be nearly stopped for appropriate values of free parameter $q$. The possibility for stopping of soliton pulses can also be realized when all parameters of nonlinear Schr\"{o}dinger equation are constant. The numerical simulations show that the stopping behavior of solitons and rectangular solitary waves can be achieved for appropriate values of free parameters characterizing these wave solutions.
\end{abstract}

\pacs{05.45.Yv, 42.65.Tg}
\maketitle

\section{Introduction}

The balanced interplay between the chromatic dispersion and self-phase
modulation nonlinearity makes possible the formation of envelope solitons in
optical fibers \cite{Hasegawa}. Such localized light wave packets can exist
in two distinct types, bright and dark solitons, which could arise in the
anomalous and normal dispersion regimes, respectively. Compared with the
dark soliton which appears as an intensity dip in an infinitely extended
constant background with a phase jump across their density minimum, the
bright soliton is a localized pulse on a zero-intensity background \cite%
{Emplit}. Importantly, the remarkable stability nature and fascinating
collision dynamics of solitons make them powerful candidates for high-speed,
high-band width and long-distance all-optical communication.

Recently, studies of soliton dynamics in inhomogeneous optical fibers have
been of considerable interest \cite{Kru1,Kru2,F1,F2,F3,F4,F5}. This because the
inhomogeneity occurs naturally in real physical media due to various factors
such as the imperfection of manufacture, fluctuation of the fiber diameters
and variation in the lattice parameters of the fiber media \cite%
{Abdullaev,Nakk,Tian}, and therefore investigations on light pulse
transmission inside inhomogeneous systems is of great interest in the area
of fiber optics communications. We note that the presence of the
inhomogeneities influences the fiber characteristic parameters such as the
self-phase modulation, group velocity dispersion, and optical fiber gain or
loss gain coefficient, which become functions of the evolution variable.
Remarkably, the variation of fiber parameters may strongly affects the
dynamical behavior, stability, and interaction of propagating solitons.

With consideration of the inhomogeneities in medium, the dynamics of 
light pulse propagation is governed by the nonlinear Schr\"{o}dinger
equation (NLSE) with coefficients being functions of time \cite{JFung}, or
equivalently, the variable representing the propagation distance \cite%
{Kru1,Kru2}. Interestingly, studying the propagation of optical waves within
the framework of such inhomogeneous equation revealed a multitude of novel
behaviors not found in the homogeneous case, for instance, by introducing
the concepts of self-similar solitons and nonautonomous solitons \cite%
{Kru1,Kru2,Kr3,Kr4,SH1,SH2,SH3,SH4,SH5}.

Currently, the optimal control of the soliton parameters also called the
soliton management has been paid much attention due to its potential
applications in the field of fiber-optic communication \cite{Belya,R1,R2}.
In this regard, impressive results have been obtained with studying the
control of the amplitude or shape of envelope solitons under nonlinearity
and dispersion managements in nonlinear systems governed by the NLS family
of equations with variable coefficients. Moreover, soliton dispersion,
amplification, and soliton pulse width managements as well as combined
nonlinear and dispersion soliton management regime were discussed in \cite%
{SH1,SH2, S2,S4}. Although the soliton management has been widely employed
for modifying properties such as the soliton's width and amplitude, its
application to modify the velocity of the soliton has not been widespread.
It is worthy to note that besides the width and amplitude, the wave velocity
can also describe the dynamical behavior of soliton evolution. Very
recently, the wave-speed management of solitons and periodic waves has been
discussed within the framework of the cubic NLSE with time-modulated
second-order dispersion and cubic nonlinearity \cite{Luke}. A challenging
problem in this context is how to control the velocity of the soliton under
the effects of spatially modulated dispersion and nonlinearity. It should be
pointed out that the soliton propagation in an optical fiber is generally
examined by treating the spatial variable as the evolution variable \cite%
{Tang}. We also note here that investigations of wave velocity management in
the framework of the generalized NLSE with spatially modulated coefficients
is significantly more complicated than the one modelled by the NLSE with
time-dependent coefficients. 

In this paper, we
consider for the first time a new (generalized) type of soliton solution for
the NLSE describing the propagation and stopping of optical pulses in fiber
waveguide with varying and constant parameters. The solution of stopping
problem of optical solitons is presented in this paper using the simplest
form of NLSE. We emphasize that the same approach for stopping solitons can
be used for much more complicated models based on nonlinear Schr\"{o}dinger
equation and also coupled nonlinear Schr\"{o}dinger equations. Thus, the
described approach leads to numerous applications for stopping problem of
optical pulses in fiber waveguides.

The rest of this paper is organized as follows. In Sec. II, we describe the
similarity transformation reducing the generalized NLSE with spatially
inhomogeneous dispersion and nonlinearity to the standard NLSE with constant
coefficients. We also present here the self-similar variables and constraint
for the waveguide parameters. Moreover, we construct the analytical
self-similar bright soliton solution of the model and determine the precise
expression of the soliton velocity. In Sec. III, we examine the existence
and properties of the generalized soliton solution of NLSE with constant
parameters. Section IV mainly presents two types of self-similar dark
soliton solutions and four types of self-similar periodic wave solutions for
the model, which demonstrate the richness of generalized
solitary and periodic waves in optical
waveguide with varying and constant parameters. In
Sec. V we present the numerical results for the stopping behavior of self-similar bright, dark and rectangular
solitons for a special choice of dispersion parameters. Finally, we give some conclusions in Sec. VI.

\section{Generalized soliton solution of NLSE with varying parameters}

In this section, we are considering the following equation with distributed
coefficients: 
\begin{equation}
i\frac{\partial\psi}{\partial z}+i\beta _{1}(z)\frac{\partial \psi}{\partial
t}- \frac{1}{2}\beta _{2}(z)\frac{\partial ^{2}\psi}{\partial t^{2}}+\gamma
(z)\left\vert \psi\right\vert ^{2}\psi=0.  \label{1}
\end{equation}

To determine the exact self-similar solutions for Eq. (\ref{1}) we start
with constructing the transformation like \cite{Dai2,Dai3,Liu}: 
\begin{equation}
\psi(z,t)=U \left(Z(z),T(z,t)\right) e^{i\theta (z,t)},  \label{2}
\end{equation}
where $U(Z,T)$ is a complex function of $Z$ and $T$. Here $T(z,t)$, $Z(z)$,
and $\theta (z,t)$ are real functions to be determined. Equations (\ref{1})
and (\ref{2}) yield the following NLS equation as 
\begin{equation}
i\frac{\partial U }{\partial Z}-\frac{\alpha }{2}\frac{\partial ^{2}U }{%
\partial T^{2}}+\sigma \left\vert U \right\vert ^{2}U=0,  \label{3}
\end{equation}%
with the following constraint relations, 
\begin{equation}
\theta _{tt}=0,~~~~T_{z}+~\beta _{1}T_{t}-\beta _{2}\theta _{t}T_{t}=0,
\label{4}
\end{equation}%
\begin{equation}
\theta _{z}+\beta _{1}\theta _{t}-\frac{\beta _{2}}{2}\theta
_{t}^{2}=0,\quad T_{tt}=0,  \label{5}
\end{equation}%
\begin{equation}
\beta _{2}T_{t}^{2}=\alpha Z_{z},\quad \gamma (z)=\sigma Z_{z},  \label{6}
\end{equation}
where $\alpha$ and $\sigma$ are some constant parameters. Solving this
set of equations with initial conditions $T=0$, $Z=0$ and $\theta=\theta_{0}$
for $t=z=0$ we find that the similarity variable $T(z,t)$, effective
propagation distance $Z(z)$ and phase $\theta (z,t)$ of the pulse are of the
form:%
\begin{equation}
T(z,t)=pt-p\int_{0}^{z}\beta _{1}(s)ds+pq\int_{0}^{z}\beta _{2}(s)ds,
\label{7}
\end{equation}%
\begin{equation}
Z(z)=\frac{p^{2}}{\alpha }\int_{0}^{z}\beta _{2}(s)ds,  \label{8}
\end{equation}%
\begin{equation}
\theta (z,t)=qt-q\int_{0}^{z}\beta _{1}(s)ds+\frac{q^{2}}{2}%
\int_{0}^{z}\beta _{2}(s)ds+\theta _{0},  \label{9}
\end{equation}
where $p$, $q$ and $\alpha$ are the arbitrary real parameters.

The second equation in (\ref{6}) ($\gamma (z)=\sigma Z_{z}$) yields the
following constraint for the waveguide parameters:%
\begin{equation}
\gamma (z)=\frac{\sigma p^{2}}{\alpha} \beta _{2}(z).  \label{10}
\end{equation}
However, we can consider the constraint (\ref{10}) at $z=0$ (or at any point 
$z=z_{0}$) which yields the parameter $\sigma$ as 
\begin{equation}
\sigma=\frac{\alpha\gamma(0)}{p^{2}\beta _{2}(0)}.  \label{11}
\end{equation}
Thus, the parameter $\sigma$ is not an arbitrary parameter in this theory
because it is given by Eq. (\ref{11}). Hence, really we have in this
approach only three nontrivial independed free parameters as $p$, $k$ and $%
\alpha$. Using Eq. (\ref{11}) we can rewrite the constraint (\ref{10}) in
the following simple and suitable form: 
\begin{equation}
\gamma (z)=\frac{\gamma (0)}{\beta _{2}(0)} \beta _{2}(z).  \label{12}
\end{equation}

We can consider the traveling solutions of Eq. (\ref{3}) in the form: 
\begin{equation}
U(Z,T)=f(\zeta )\exp [i(\kappa Z-\delta T)],  \label{13}
\end{equation}%
where $f(\zeta )$ is a real function of variable $\zeta =T-\epsilon Z$. The
substitution of Eq. (\ref{13}) to Eq. (\ref{3}) yields relation $\epsilon
=\alpha \delta $ and the following equation for the function $f(\zeta )$: 
\begin{equation}
\frac{d^{2}f}{d\zeta ^{2}}-af+bf^{3}=0,  \label{14}
\end{equation}%
where the parameters $a$ and $b$ are given by%
\begin{equation}
a=\delta ^{2}-\frac{2\kappa }{\alpha },\quad b=-\frac{2\sigma }{\alpha }=-%
\frac{2\gamma (0)}{p^{2}\beta _{2}(0)}.  \label{15}
\end{equation}%
We note that one can consider the coefficient $a$ as free parameter because $%
\delta $, $\kappa $ and $\alpha $ are arbitrary parameters. Fixing the
parameter $a$ we have the relation for parameters $\delta $, $\kappa $ and $%
\alpha $ given in Eq. (\ref{15}).

We can write the variable $\zeta=T-\alpha\delta Z$ in the form $%
\zeta=p\tau(z,t)$ where $\tau(z,t)$ is given by 
\begin{equation}
\tau(z,t)=t-\int_{0}^{z}\beta _{1}(s)ds+(q-\delta p)\int_{0}^{z}\beta
_{2}(s)ds,  \label{16}
\end{equation}%
with $\tau=0$ for $t=0$ and $z=0$. The soliton solution of elliptic Eq. (\ref{14}) is 
\begin{equation}
f(\zeta)=\pm\left(\frac{2a}{b}\right)^{1/2}\mathrm{sech}(\sqrt{a}%
(\zeta-\zeta_{0})),  \label{17}
\end{equation}
with $\zeta=p\tau(z,t)$ and $\zeta_{0}$ is an arbitrary real parameter
(integration constant). Thus, the soliton solution of Eq. (\ref{1}) with
varying parameters is given by 
\begin{equation}
\psi(z,t)=\pm\left(\frac{2a}{b}\right)^{1/2}\mathrm{sech} (w(\tau(z,t)-%
\tau_{0})) \exp \left[i\Phi(z,t)\right], ~~~~\Phi(z,t)=\kappa Z(z)-\delta
T(z,t)+\theta(z,t),  \label{18}
\end{equation}%
where $w=p\sqrt{a}$ and $|w|$ is the inverse width of soliton and $%
\tau_{0}=\zeta_{0}/p$. The velocity of this soliton pulse follows from
equation, 
\begin{equation}
d\tau=dt+(-\beta _{1}(z)+(q-\delta p)\beta _{2}(z))dz=0.  \label{19}
\end{equation}%
Thus, the velocity of soliton is $\mathrm{v}(z)=dz/dt$: 
\begin{equation}
\mathrm{v}(z)=\frac{1}{\beta _{1}(z)+(\delta p-q)\beta _{2}(z)},  \label{20}
\end{equation}
We can define the functions $\beta _{2}(z)$ and $\gamma(z)$ in the form that
the constraint relation (\ref{12}) is satisfied. The soliton solution given
by Eq. (\ref{18}) depends on five arbitrary parameters $\delta$, $\kappa$, $%
p $, $q$, $\alpha$ and also two arbitrary integration constants $\tau_{0}$
and $\theta _{0}$.

However, the optical pulse is assumed to be quasi-monochromatic, and hence
the width $\tau_{0}=1/|w|$ of the pulses is restricted by condition $%
\tau_{0}\omega_{0}\gg 1$ where $\omega_{0}$ is the frequency at which the
pulse spectrum is centered. It follows from Eq. (\ref{20}) that the velocity 
$\mathrm{v}(z)$ of soliton can be less any given small value $\mathrm{v}_{0}$
for appropriate free parameters $q$ because the velocity $\mathrm{v}(z)$
tends to zero when the free parameter $q$ tends to infinity. Thus, the
solitons describing by Eq. (\ref{18}) can be nearly stopped for appropriate
values of free parameter $q$.

\section{Generalized soliton solution of NLSE with constant parameters}

In this section, we are considering Eq. (\ref{1}) with constant coefficients 
$\beta _{1}$, $\beta _{2}$ and $\gamma$ \cite{Agrawal}: 
\begin{equation}
i\psi_{z}+i\beta _{1}\psi_{t}- \frac{1}{2}\beta
_{2}\psi_{tt}+\gamma\left\vert \psi\right\vert ^{2}\psi=0.  \label{21}
\end{equation}%
In this case the solution of Eqs. (\ref{4})-(\ref{6}) yields 
\begin{equation}
T(z,t)=pt-p\beta _{1}z+pq\beta _{2}z,  \label{22}
\end{equation}%
\begin{equation}
Z(z)=\frac{p^{2}}{\alpha }\beta _{2}z,  \label{23}
\end{equation}%
\begin{equation}
\theta (z,t)=qt-q\beta _{1}z+\frac{q^{2}}{2} \beta _{2}z+\theta _{0},
\label{24}
\end{equation}
where $p$, $q$ and $\alpha$ are the arbitrary real parameters. The parameter 
$\sigma$ follows from (\ref{11}) as 
\begin{equation}
\sigma=\frac{\alpha\gamma}{p^{2}\beta _{2}}.  \label{25}
\end{equation}
One can see that the constraint (\ref{12}) is always satisfied when the
parameters $\beta _{2}$ and $\gamma$ are constants. The parameters $a$ and $%
b $ for constant $\beta _{2}$ and $\gamma$ follow from Eq. (\ref{15}) as 
\begin{equation}
a=\delta^{2}-\frac{2\kappa}{\alpha},\quad b=-\frac{2\gamma}{p^{2}\beta _{2}}.
\label{26}
\end{equation}
The variable $\zeta=T-\alpha\delta Z$ is given in this case by (\ref{16}) in
the following form: 
\begin{equation}
\zeta(z,t)=p\tau(z,t),\quad \tau(z,t)=t-[\beta_{1}+(\delta p-q)\beta_{2}]z.
\label{27}
\end{equation}%
Thus, the soliton solution follows from Eq. (\ref{18}) as 
\begin{equation}
\psi(z,t)=\pm\left(\frac{2a}{b}\right)^{1/2}\mathrm{sech} (w(t-\mathrm{v}%
^{-1}z-\tau_{0})) \exp \left[i\Phi(z,t)\right],  \label{28}
\end{equation}
where $w=p\sqrt{a}$ and $|w|$ is the inverse width of soliton, and the
velocity $\mathrm{v}$ of soliton is 
\begin{equation}
\mathrm{v}=\frac{1}{\beta _{1}+(\delta p-q)\beta _{2}}.  \label{29}
\end{equation}
The phase $\Phi(z,t)$ of this soliton solution is given by $\Phi(z,t)=\kappa
Z(z)-\delta T(z,t)+\theta(z,t)$ or in explicit form as 
\begin{equation}
\Phi(z,t)=(q-p\delta)t+(p\delta-q)\beta _{1}z+\left(\frac{\kappa p^{2}}{%
\alpha}+\frac{q^{2}}{2}-pq\delta\right)\beta_{2}z+\theta_{0}.  \label{30}
\end{equation}%
This new type soliton solution exists when $a>0$ and $b>0$. The soliton
solution given by Eq. (\ref{28}) depends on five arbitrary parameters: $%
\delta$, $\kappa$, $p$, $q$, $\alpha$ and integration constants $\tau_{0}$
and $\theta _{0}$. The width $\tau_{0}=1/|w|$ of pulses is restricted by
condition $\tau_{0}\omega_{0}\gg 1$. Since we have $\omega_{0}\simeq 10^{15}
s^{-1}$ the width of pulses satisfy the condition $\tau_{0}\gtrsim 0.1 ps$.
It follows from Eqs. (\ref{26}) and (\ref{29}) that the velocity $\mathrm{v}$
is positive and tends to zero when $\beta _{2}<0$, $\gamma>0$ and $%
q\rightarrow +\infty$; and the velocity $\mathrm{v}$ is positive and tends
to zero when $\beta _{2}>0$, $\gamma<0$ and $q\rightarrow -\infty$.

We note that in the particular case when free parameters $p$, $q$, and $%
\alpha$ are given as $p=1$, $q=0$, $\alpha=\beta _{2}$ we have the variables
given by Eqs. (\ref{22})-(\ref{24}) in the form: $T=t-\beta _{1}z$, $Z=z$
and $\theta=\theta_{0}$. In this case the solution in Eq. (\ref{28}) leads
to the standard form for soliton solution: 
\begin{equation}
\psi(z,t)=\pm\left(\frac{2a}{b}\right)^{1/2}\mathrm{sech}(\sqrt{a}(t-\mathrm{%
v}^{-1}z-\tau_{0})) \exp \left[i(\tilde{\kappa}z-\delta t+\theta_{0})\right],
\label{31}
\end{equation}%
where $\tilde{\kappa}=\kappa+\delta\beta _{1}$, $a=\delta^{2}-2\kappa/\beta
_{2}$ and $b=-2\gamma/\beta _{2}$. The velocity of this soliton follows from
Eq. (\ref{29}) in the form: $\mathrm{v}=1/(\beta _{1}+\delta\beta _{2})$.
This means that such sort of solitons can not be stopped because the the
physical perimeters $\beta _{1}$ and $\beta _{2}$ are fixed in the optical
waveguide.

\section{Solitary and periodic generalized wave solutions}

In this section we present a set of solitary and periodic self-similar wave
solutions based on results obtained in Sec. II. All these wave solutions
depend on five arbitrary parameters $\delta$, $\kappa$, $p$, $q$, $\alpha$
and two arbitrary integration constants $\tau_{0}$ and $\theta _{0}$. The
parameters $a$ and $b$ of these solutions are given by Eq. (\ref{15}) and
the phase is given by $\Phi(z,t)=\kappa Z(z)-\delta T(z,t)+\theta(z,t)$ or
in explicit form as 
\begin{equation}
\Phi(z,t)=(q-p\delta)t+(p\delta-q)\int_{0}^{z}\beta _{1}(s)ds+\left(\frac{%
\kappa p^{2}}{\alpha}+\frac{q^{2}}{2}-pq\delta \right)\int_{0}^{z}\beta
_{2}(s)ds+\theta_{0}.  \label{32}
\end{equation}%
The function $\tau(z,t)$ and velocity $\mathrm{v}(z)$ of these self-similar
wave solutions are given in Eqs. (\ref{16}) and (\ref{20}). Moreover, we
consider here the functions $\beta _{2}(z)$ and $\gamma(z)$ in the form that
constraint (\ref{12}) is satisfied. Note that in the case when coefficients $%
\beta _{1}$, $\beta _{2}$ and $\gamma$ in Eq. (\ref{1}) are constants the
constraint given in Eq. (\ref{12}) is satisfied automatically and the
function $\tau(z,t)$ and velocity $\mathrm{v}$ are given by Eqs. (\ref{27})
and (\ref{29}). We also shown above that wave solutions given in Eqs. (\ref%
{18}) and (\ref{28}) can be stopped when free parameter $q$ tends to
infinity. This means that the velocity $\mathrm{v}(z)$ of solitary pulses
can be less any given small value $\mathrm{v}_{0}$ for appropriate free
parameters $q$. The bright soliton solution for varying and constant
parameters of Eq. (\ref{1}) are given in Eqs. (\ref{18}) and (\ref{28}). The
other solitary and periodic solutions of Eq. (\ref{1}) we present below in
this section. Note that all these solutions can also be stopped when free
parameter $q$ tends to infinity.

\begin{description}
\item[\textbf{Kink solitary waves for $a<0$ and $b<0$}] 
\end{description}

The solution of Eq. (\ref{14}) leads to the kink wave solution as%
\begin{equation}
f(\zeta )=\pm \Lambda \mathrm{tanh}(w_{0}(\zeta -\zeta_{0})).  \label{33}
\end{equation}%
The parameters of this solution are%
\begin{equation}
\Lambda=\sqrt{\frac{a}{b}},~~~~w_{0}=\sqrt{\frac{-a}{2}},  \label{34}
\end{equation}%
where $a<0$ and $b<0$. Hence, the kink solitary solution for Eq. (\ref{1})
is 
\begin{equation}
\psi (z,t)=\pm \Lambda \mathrm{tanh}(w(\tau(z,t)-\tau_{0}))\exp[i\Phi(z,t)],
\label{35}
\end{equation}%
where $w=pw_{0}$. Note that this kink solution has the form of dark soliton
for intensity $I=|\psi (z,\tau )|^{2}=\Lambda^{2}\mathrm{tanh}%
^{2}(w(\tau(z,t)-\tau_{0}))$.

\begin{description}
\item[\textbf{Dark rational-solitary waves for $a<0$ and $b<0$}] 
\end{description}

The solution of Eq. (\ref{14}) leads to a rational-solitary wave of the
form, 
\begin{equation}
f(\zeta )=\pm \frac{A\mathrm{tanh}\left( w_{0}(\zeta -\zeta_{0})\right) }{1+%
\mathrm{sech}\left( w_{0}(\zeta -\zeta_{0})\right) }.  \label{36}
\end{equation}%
The parameters for this solitary wave are 
\begin{equation}
A=\sqrt{\frac{a}{b}},\qquad w_{0}=\sqrt{-2a},\qquad  \label{37}
\end{equation}%
where $a <0$ and $b<0$. Hence the rational-solitary wave solution for Eq. (%
\ref{1}) is given by 
\begin{equation}
\psi (z,t)=\pm \frac{A\mathrm{tanh}\left( w(\tau(z,t)-\tau_{0})\right) }{1+%
\mathrm{sech}\left( w(\tau(z,t)-\tau_{0})\right) }\exp[i\Phi(z,t)],
\label{38}
\end{equation}%
where $w=pw_{0}$. This solitary wave has the form of dark soliton for
intensity $I=|\psi (z,\tau )|^{2}$. This functional form of the solitary
wave differs from the dark solitary $\text{tanh}$-wave.

\begin{description}
\item[\textbf{Periodic waves for $a>0$ and $b>0 $}] 
\end{description}

We have found the periodic solution of Eq. (\ref{14}) as 
\begin{equation}
f(\zeta )=\pm [A+B\mathrm{cn^{2}}(w_{0}(\zeta -\zeta_{0}),k)]^{1/2},~~~~~~~
\label{39}
\end{equation}%
where the modulus $k$ of Jacobi elliptic function $\mathrm{cn}(w_{0}(\zeta
-\zeta_{0},k)$ is given in the interval $0<k<1$. The parameters of periodic
solution are 
\begin{equation}
A=\frac{2a(1-k^{2})}{b(2-k^{2})},~~~~B=\frac{2a k^{2}}{b(2-k^{2})},~~~~~~~
\label{40}
\end{equation}%
\begin{equation}
w_{0}=\sqrt{\frac{a}{2-k^{2}}}.~~~~~~~  \label{41}
\end{equation}
It follows from this solution the conditions for parameters as $a>0$ and $%
b>0 $. Thus, the wave function (\ref{39}) yields the family of periodic wave
solutions for the NLS equation (\ref{1}) as 
\begin{equation}
\psi (z,t)=\pm \lbrack A+B\mathrm{cn^{2}}(w(\tau(z,t)-\tau_{0}),k)]^{1/2}%
\exp [i\Phi(z,t)],  \label{42}
\end{equation}%
where $w=pw_{0}$ and the modulus $k$ is an arbitrary parameter in the
interval $0<k<1$ and $\tau_{0}$ is an arbitrary real constant. We note that
in the limiting cases with $k=1$ this periodic wave reduces to bright-type
soliton solution given in Eq. (\ref{18}).

\begin{description}
\item[\textbf{Periodic waves for $a<0$, $b>0$ and $a>0$, $b>0$}] 
\end{description}

In the case when parameters belong these intervals we have found the
periodic solution of Eq. (\ref{14}) as 
\begin{equation}
f(\zeta )=\pm \Lambda \mathrm{cn}(w_{0}(\zeta -\zeta_{0}),k),  \label{43}
\end{equation}%
where $k$ is the modulus of Jacobi elliptic function $\mathrm{cn}%
(w_{0}(\zeta -\zeta_{0}),k)$. The parameters $\Lambda $ and $w$ are given as 
\begin{equation}
\Lambda=\sqrt{\frac{2ak^{2}}{b(2k^{2}-1)}},~~~~ w_{0}=\sqrt{\frac{a}{2k^{2}-1%
}}.  \label{44}
\end{equation}%
In this solution the modulus $k$ can belong two different intervals: $0<k<1/%
\sqrt{2}$ or $1/\sqrt{2}<k<1$. It follows from this solution the conditions
for parameters as $a<0$ and $b>0$ when $0<k<1/\sqrt{2}$, and the conditions
for parameters are $a>0$ and $b>0$ when $1/\sqrt{2}<k<1 $. The solution (\ref{43}) yields the following family of periodic wave solutions for the
generalized NLSE (\ref{1}) as 
\begin{equation}
\psi (z,t)=\pm \Lambda\mathrm{cn}(w(\tau(z,t)-\tau_{0}),k)\exp [i\Phi(z,t)],
\label{45}
\end{equation}%
where $w=pw_{0}$ and the modulus $k$ is an arbitrary parameter in the
interval $0<k<1/\sqrt{2}$ or $1/\sqrt{2}<k<1$. In the limiting case with $%
k=1 $ this solution reduces to bright-type soliton solution given in Eq. (%
\ref{18}).

\begin{description}
\item[\textbf{Periodic waves for $a<0$ and $b<0 $}] 
\end{description}

In the case when parameters of wave solution belong the intervals $a<0$ and $%
b<0$ we have found that Eq. (\ref{14}) yields the periodic solution as 
\begin{equation}
f(\zeta )=\pm \Lambda \mathrm{sn}(w_{0}(\zeta -\zeta_{0}),k),  \label{46}
\end{equation}%
where $0<k<1$. Here $\mathrm{sn}(w_{0}(\zeta -\zeta_{0}),k)$ is the Jacobi
elliptic function with modulus $k$. The parameters $\Lambda $ and $w$ are
given by%
\begin{equation}
\Lambda=\sqrt{\frac{2ak^{2}}{b(1+k^{2})}},~~~~w_{0}=\sqrt{\frac{-a}{1+k^{2}}}%
.  \label{47}
\end{equation}%
It follows from this solution the conditions for parameters as $a<0$ and $%
b<0 $. The solution (\ref{46}) yields the following family of periodic wave
solutions for the generalized NLSE (\ref{1}) as 
\begin{equation}
\psi (z,t )=\pm \Lambda \mathrm{sn}(w(\tau(z,t)-\tau_{0}),k)\exp
[i\Phi(z,t)],  \label{48}
\end{equation}%
where $w=pw_{0}$ and the modulus $k$ is an arbitrary parameter in the
interval $0<k<1$. In the limiting case with $k=1$ this solution reduces to
the kink wave solution given in Eq. (\ref{35}).

\begin{description}
\item[\textbf{Periodic rational-elliptic waves for $a<0$ and $b<0$}] 
\end{description}

In the case when parameters of wave solution belong this intervals we have
found that Eq. (\ref{14}) yields the periodic rational-elliptic solution as 
\begin{equation}
f(\zeta )=\pm \frac{A\mathrm{sn}\left( w_{0}(\zeta -\zeta _{0}),k\right) }{1+%
\mathrm{dn}\left( w_{0}(\zeta -\zeta _{0}),k\right) },  \label{49}
\end{equation}%
where $0<k<1$. The parameters $A$ and $w_{0}$ for this periodic solution are%
\begin{equation}
A=\sqrt{\frac{ak^{4}}{b(2-k^{2})}},~~~~w_{0}=\sqrt{\frac{-2a}{2-k^{2}}},
\label{50}
\end{equation}%
where $a<0$ and $b<0$. Thus, Eq. (\ref{49}) yields the periodic bounded
solution of Eq. (\ref{1}) as%
\begin{equation}
\psi (z,t)=\pm \frac{A\mathrm{sn}(w(\tau (z,t)-\tau _{0}),k)}{1+\mathrm{dn}%
(w(\tau (z,t)-\tau _{0}),k)}\exp [i\Phi (z,t)],  \label{51}
\end{equation}%
where $w=pw_{0}$ and the modulus $k$ is an arbitrary parameter in the
interval $0<k<1$. One notes that in the limiting case with $k=1$ this
solution reduces to the dark rational-solitary waves given in Eq. (\ref{38}).

\section{Numerical results for generalized bright, dark and rectangular solitons}

In this section, we investigate the stopping behavior of the obtained
self-similar bright and dark solitons by direct numerical simulation for a special controlled soliton system. We utilize the split-step Fourier method \cite{Agrawal} for numerical integration of Eq. (\ref{1}). 
We first consider the case of an optical fiber with constant values of $\beta _{1}$, $\beta _{2}$ and $\gamma $. Figures 1(a) and 1(b) show the
evolution of bright and dark solitons inside the optical fiber system with initial wave function given in Eqs. (\ref{28}) and (\ref{35}). Note that the kink solution in (\ref{35}) has the form of dark soliton
for intensity $I=|\psi (z,\tau )|^{2}=\Lambda^{2}\mathrm{tanh}%
^{2}(w(\tau(z,t)-\tau_{0}))$.
As seen in Fig. 1, the soliton pulses propagate with the invariant
amplitude and width in such a situation. Consider now an
inhomogeneous fiber with exponentially varying parameters:%
\begin{equation}
\beta _{1}(z)=d_{1}\exp (-gz),~~~~\beta _{2}(z)=d_{2}\exp (-hz),  \label{52}
\end{equation}
\noindent where $d_{1},$ $d_{2},$ $g$ and $h$ are real constants. The nonlinearity parameter $\gamma (z)$ can be obtained by Eq. (\ref{12}).

\begin{figure}[h]
\includegraphics[width=1\textwidth]{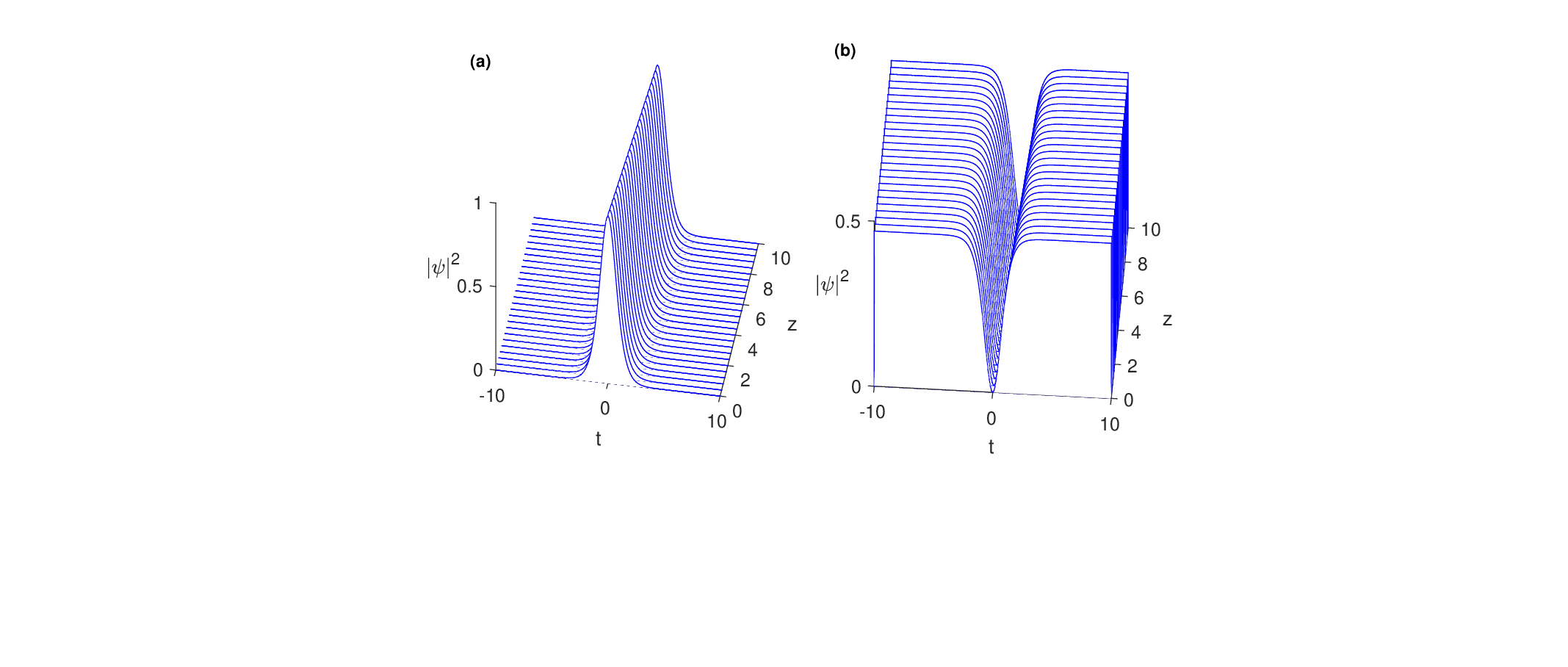}
\caption{Numerical evolutions of (a) the bright soliton when $\protect\beta%
_{1}(0)=-0.1,$ $\protect\beta _{2}(0)=-1,$ $\protect\gamma (0)=1$ and (b)
dark soliton when $\protect\beta _{1}(0)=0.194,$ $\protect\beta %
_{2}(0)=0.47, $ $\protect\gamma (0)=1$ inside a fiber within the framework of Eq. (\ref{1}). The other parameters are $\protect\delta =-0.2,$ $\protect\kappa =0.48,$ $p=1,$ $q=0,$ and $\protect\tau _{0}=0.$}
\label{FIG.1.}
\end{figure}

\begin{figure}[h]
\includegraphics[width=1\textwidth]{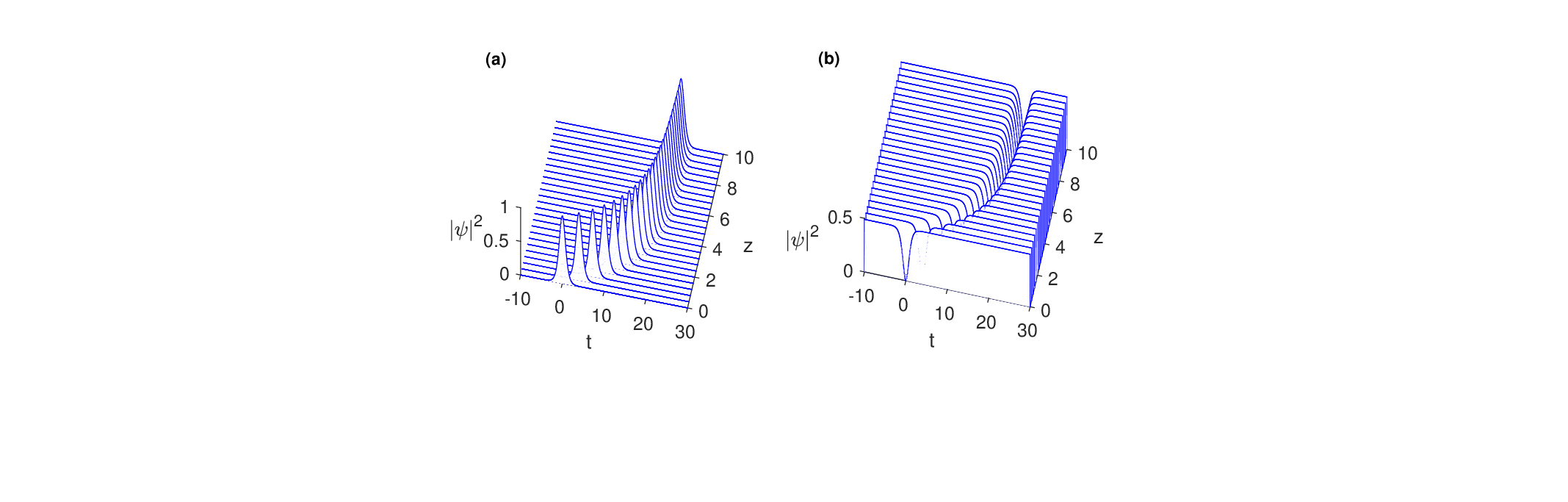}
\caption{Numerical evolutions of optical solitons with parameters given in Eq. (\ref{52}): $d_{1}=10.2,$ $d_{2}=-0.2,$ $g=0.5,$ $h=0.5,$ $\protect%
\delta =-0.5,$ $p=1,$ $q=10,$ $\protect\tau _{0}=0\ $\ (a) the bright
soliton when $\protect\alpha =-1,$ $\protect\sigma =1,$ $\protect\kappa %
=0.375$ and (b) dark soliton when $\protect\alpha =1,$ $\protect\sigma =2,$ $%
\protect\kappa =1.125.$}
\label{FIG.2.}
\end{figure}

\begin{figure}[h]
\includegraphics[width=1\textwidth]{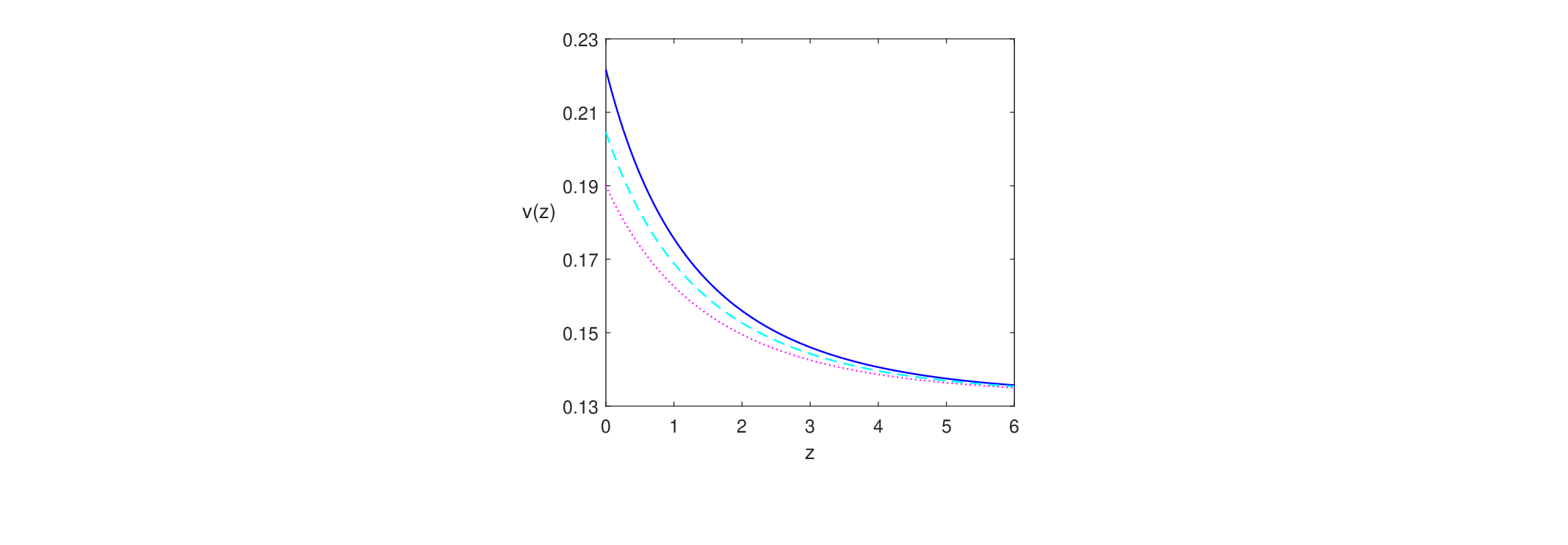}
\caption{Soliton velocity $\mathrm{v}(z)$ versus distance for different
values of free parameter $q$, $q=0.5$ (thick line), $q=0.7$ (dashed line)
and $q=0.9$ (dotted line). The other parameters are $d_{1}=2,$ $d_{2}=-0.5,$ 
$g=0,$ $h=0.5,$ $\protect\delta =2.1,$ and $p=1.$ }
\label{FIG.3.}
\end{figure}

\begin{figure}[h]
\includegraphics[width=1\textwidth]{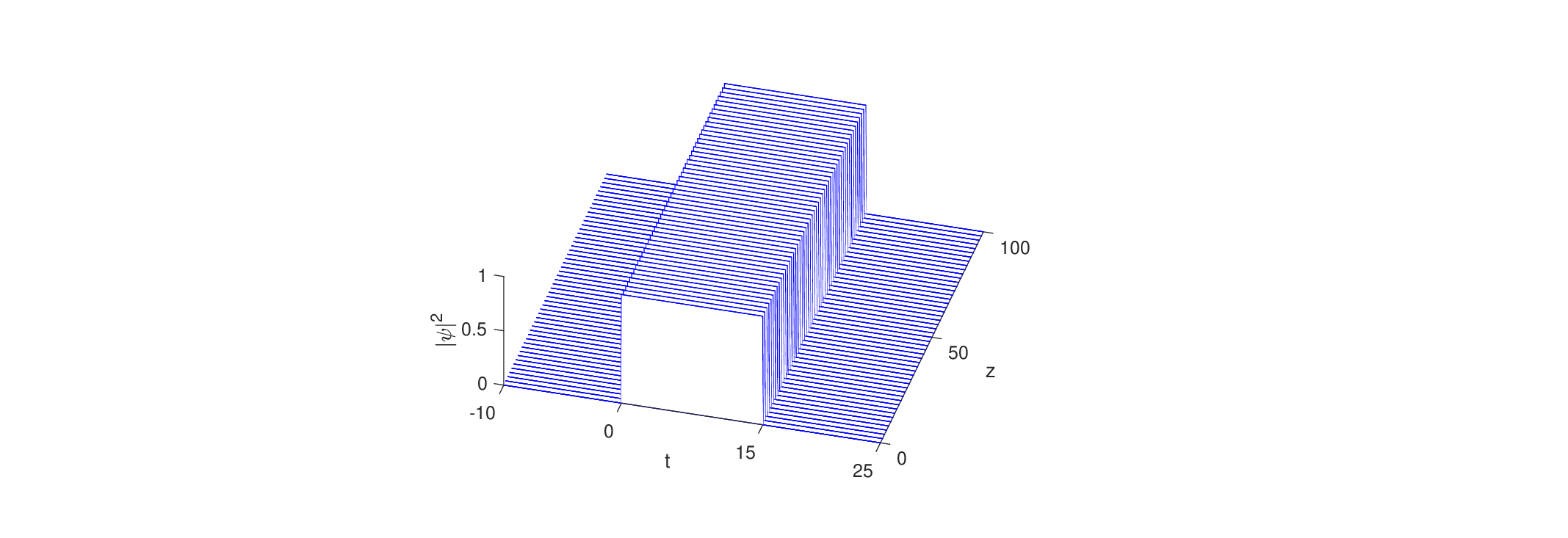}
\caption{Numerical evolution of the rectangular pulse when $\protect\beta %
_{1}=1,$ $\protect\beta _{2}=-3.54,$ $\protect\gamma =0.0625$ inside a fiber
within the framework of Eq. (\ref{1}). The other parameters are $%
\protect\delta =0,$ $\protect\kappa =0.5,$ $p=0.188,$ $q=10^{4},$ $\protect%
\alpha =-1,$ $\protect\sigma =0.5,$ and $\protect\tau _{0}=0.$}
\label{FIG.4.}
\end{figure}

\begin{figure}[h]
\includegraphics[width=1\textwidth]{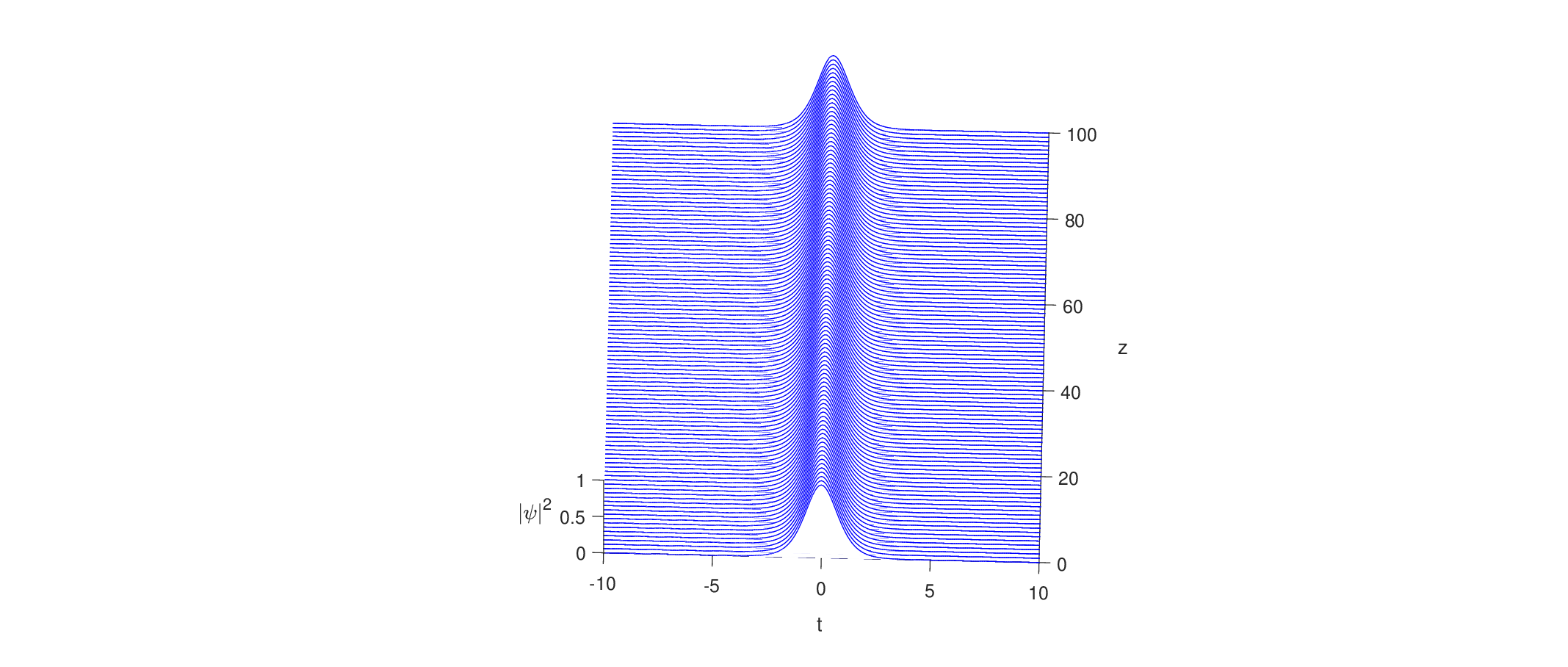}
\caption{Numerical evolution of the bright soliton when $\protect\beta %
_{1}=0.1,$ $\protect\beta _{2}=-1,$ $\protect\gamma =1$ inside a fiber
within the framework of Eq. (\ref{1}). The other parameters are $%
q=10^{4},$ $\protect\alpha =-0.5,$ $\protect\sigma =0.5,$\ $\protect\kappa %
=0.25,$\ $p=1,$ and\ $\protect\delta =0.$}
\label{FIG.5.}
\end{figure}
We present the corresponding numerical results in Fig. 2. From this figure,
in the beginning of the evolution, the self-similar bright and dark solitons
accelerate. As the solitons propagate along the fiber, they abruptly
decelerate and hence their evolution becomes slow. Under this circumstance,
if one further extend the propagation distance, the solitons can stop
steadily and have application value in the communication.

In order to strictly answer the question of stopping motion of the envelope
solitons, we have numerically computed the velocity of the solitons as a
function of the propagation distance in the fiber for three different values
of the free parameter $q$. The numerical results are presented in Fig. 3,
which show that the soliton velocity decreases dramatically in the magnitude
with the increase of transmission distance $z$. From this figure, one can
also observe that the soliton velocity has large magnitude for small values
of $q$, in full agreement with the analytical result (\ref{20}). The
decrease of velocity to zero value at large distance unambiguously confirms
the occurrence of stopping motion for solitons in the fiber.

We note that the pulses with rectangular profiles are also satisfy to Eq. (\ref{1}) except two extreme points. The numerical simulations show that such
pulses are really the solutions with a high accuracy. Such rectangular
pulses have the form $f(\zeta)=A\eta(\xi)$ where $\xi=\nu_{0}(\zeta-%
\zeta_{0}) $ and $\nu_{0}$ is a free parameter here. The function $\eta(\xi)$
is defined as $\eta(\xi)=1$ for $0\leq \xi\leq \xi_{0}$ and $\eta(\xi)=0$
otherwise. The substitution of this rectangular function to Eq. (\ref{14})
yields $A^{2}=a/b$ because $f^{\prime}=f^{\prime\prime}=0$ except two
points: $\xi=0$ and $\xi=\xi_{0}$. Thus, the approach presented in Sec. II
yields the rectangular solitary solution of NLSE (\ref{1}) with varying
parameters as 
\begin{equation}
\psi(z,t)=\pm\left(\frac{a}{b}\right)^{1/2}\eta(\nu(\tau (z,t)-\tau
_{0}))\exp [i\Phi (z,t)],  \label{53}
\end{equation}%
where $\nu=p\nu_{0}$. The function $\tau (z,t)$ and phase $\Phi (z,t)$ are
given by Eqs. (\ref{16}) and (\ref{32}). This solution reduces to the
following rectangular pulse when the parameters in NLSE (\ref{1}) are
constant: 
\begin{equation}
\psi(z,t)=\pm\left(\frac{a}{b}\right)^{1/2} \eta(\nu(t-\mathrm{v}%
^{-1}z-\tau_{0}))\exp [i\Phi (z,t)],  \label{54}
\end{equation}%
where the velocity $\mathrm{v}$ and phase $\Phi (z,t)$ are given by Eqs. (\ref{29}) and (\ref{30}).

We consider below the stopping of rectangular and soliton optical pulses
when all parameters in NLSE (\ref{1}) are constant. The approach developed
in this paper allows to prepare the slow optical pulses with an arbitrary
small velocity given by Eq. (\ref{29}). First we consider numerically the
initial pulse with rectangular shape. Let us for example $\delta=0$ and $%
\theta_{0}=0$ then the phase given in (\ref{30}) is $\Phi(0,t)=qt$. Thus,
the initial rectangular pulse at $z=0$ is given by 
\begin{equation}
\psi_{0}(t)=\left(\frac{a}{b}\right)^{1/2}\eta(\nu t)\exp (iqt),  \label{55}
\end{equation}
with $\tau_{0}=0$. The velocity of rectangular pulses for $\delta=0$ is $%
\mathrm{v}=1/(\beta _{1}-q\beta _{2})$.

Note that when the variable $\xi=\nu t$ belongs to the interval $0\leq \nu
t\leq \xi_{0}$ we have $\eta(\nu t)=1$ and $\eta(\nu t)=0$ otherwise. We
define the parameter $t_{0}$ by relation $\nu t_{0}=\xi_{0}$ then $\eta(\nu
t)=1$ in the interval $0\leq t\leq t_{0}$ and $\eta(\nu t)=0$ otherwise. We
can also define $\xi_{0}=1$ which yields the inverse width as $\nu=1/t_{0}$.

Let us consider the case when $\beta_{1}>0$, $\gamma>0$, $\beta_{2}<0$ and $%
q\gg 1$ then appropriate numerical results are presented in Fig. 4 which
demonstrate that the propagating pulse has velocity almost zero. Note that
for large values of free parameter $q$ the velocity of pulses is very close
to zero. Moreover, the velocity of such optical pulses tends to zero for
enough large values of $q$.

We also consider numerically in Fig. 5 the propagation of soliton presented
in Eq. (\ref{28}) with $\delta =0$ and initial wave function at $z=0$ as 
\begin{equation}
\psi _{0}(t)=\left( \frac{2a}{b}\right) ^{1/2}\mathrm{sech}(wt)\exp (iqt),
\label{56}
\end{equation}%
with $w=p\sqrt{a}$. The Fig. 5 demonstrates that the propagating soliton
has velocity very close to zero and this velocity tends to zero for enough
large values of $q$ which follows from Eq. (\ref{29}).

The results presented above showed for the first time that the stopping
process of soliton pulses is possible in the nonlinear Kerr medium by
choosing the parameters $\beta _{1}$ and $\beta _{2}$ of the fiber material
and the free parameter $q$ appropriately. These findings may help in
realizing of the stopped solitons experimentally not only in optical fibers,
but also in planar waveguides and Bose--Einstein condensates for which the
NLSE is the key model for describing the wave dynamics. No doubt, such
stopping motion can be of practical interest especially in the area of fiber
optics communications.

\section{Conclusion}

In conclusion, we have presented a principally new approach to the problem
of the soliton stopping motion in  optical fiber medium with
a Kerr nonlinear response. Considering the generalized nonlinear Schr\"{o}dinger equation with a term resulting from the group velocity and spatially
distributed coefficients, we have derived the self-similar bright, dark and rectangular
solitons as well as self-similar periodic wave solutions of the model by
employing the similarity transformation method. All these generalized wave solutions depend on five arbitrary parameters and two free integration constants. We have also determined the
precise expression of soliton's velocity, which shows that the velocity is
related to the varying group velocity parameter $\beta _{1}(z)$,
second-order dispersion function $\beta _{2}(z)$, angular frequency $\delta $%
, and real constants $p$ and $q$. A remarkable result is that the velocity
of the obtained self-similar solitons is inversely proportional to a free
parameters $q$, which shows that the soliton's velocity can be less any
given small value $\mathrm{v}_{0}$ for appropriate free parameters $q$. This
because the velocity of solitons approaches zero when the free parameter $q$
approaches infinity, thus indicating that the solitons can be nearly stopped
for appropriate values of free parameter $q$. The numerical results show
that the stopping behavior of solitons can be readily achieved when
considering the case of dispersion decreasing fibers. The approach described in this paper leads to numerous significant applications for stopping problem of optical pulses in fiber waveguides. The stopping motion of solitons can be of practical interest especially in the area of fiber optics communications.


\begin{thebibliography}{99}
\bibitem{Hasegawa} A. Hasegawa and F. Tappert, Appl. Phys. Lett. 23, \textbf{%
142} (1973); 23, \textbf{171} (1973).

\bibitem{Emplit} P. Emplit, J. P. Hamaide, F. Reinaud, C. Froehly, and A.
Bartelemy, Opt. Commun. \textbf{62}, 374-379 (1987).

\bibitem{Kru1} V. I. Kruglov, A. C. Peacock, and J. D. Harvey, Phys. Rev.
Lett. \textbf{90}, 113902 (2003).

\bibitem{Kru2} V. I. Kruglov, A. C. Peacock, and J. D. Harvey, Phys. Rev. E 
\textbf{71}, 056619 (2005).

\bibitem{F1} Y. Kubota and T. Odagaki, Phys. Rev. E \textbf{68}, 026603
(2003).

\bibitem{F2} W. J. Liu, B. Tian, and H. Q. Zhang, Phys. Rev. E \textbf{78},
066613 (2008).

\bibitem{F3} H. Q. Zhang, B. G. Zhai1 and X. L. Wang, Phys. Scr. \textbf{85}%
, 015006 (2012)

\bibitem{F4} H. Triki and V. I. Kruglov, Phys. Rev. E \textbf{101}, 042220
(2020).

\bibitem{F5} V. I. Kruglov and H. Triki, Phys. Rev. A \textbf{103}, 013521
(2021).

\bibitem{Abdullaev} F. Abdullaev, S. Darmanyan, and P. Khabibullaev, Optical
Solitons (Springer-Verlag, Berlin, 1991).

\bibitem{Nakk} K. Nakkeeran, Phys. Rev. E \textbf{62}, 1313 (2000).

\bibitem{Tian} W.-J. Liu, B. Tian, and H.-Q. Zhang, Phys. Rev. E \textbf{78}%
, 066613 (2008).

\bibitem{JFung} J.F. Zhang, C.Q. Dai, Q. Yang, J.M. Zhu, Opt. Commun. 
\textbf{252,} 408-421\ (2005).

\bibitem{Kr3} V. I. Kruglov and H. Triki, Phys. Rev. A \textbf{102}, 043509 (2020).

\bibitem{Kr4} H. Triki and V. I. Kruglov, Chaos, Sol. and Frac. \textbf{143}, 110551
(2021).

\bibitem{SH1} V. N. Serkin, A. Hasegawa, Phys. Rev. Lett. \textbf{85},
4502-4505 (2000).

\bibitem{SH2} V. N. Serkin, A. Hasegawa, IEEE J. Sel. Top. Quant. Electron. 
\textbf{8}, 418-431 (2002).

\bibitem{SH3} S. Chen and L. Yi, Phys. Rev. E \textbf{71,} 016606 (2005)

\bibitem{SH4} S. A. Ponomarenko and G. P. Agrawal, Phys. Rev. Lett. \textbf{%
97}, 013901 (2006)

\bibitem{SH5} V. N. Serkin, A. Hasegawa, and T. L. Belyaeva, Phys. Rev.
Lett. \textbf{98}, 074102 (2007).

\bibitem{Belya} V. N. Serkin, T. L. Belyaeva, Quantum Electronics \textbf{31}
(11) 1007-1015 (2001).

\bibitem{R1} W. Snyder and D. J. Mitchell, Opt. Lett. \textbf{18}, 101-103
(1993).

\bibitem{R2} M. Saha, A. K. Sarma and A. Biswas, Phys. Lett. A \textbf{373},
4438-4441 (2009).

\bibitem{S2} V. N. Serkin, A. Hasegawa, JETP Lett. \textbf{72}, 89-92 (2000).

\bibitem{S4} V. N. Serkin, T. L. Belyaeva, JETP Lett. \textbf{74}, 573--577
(2001).

\bibitem{Luke} L. W. S. Baines and . A. Van Gorder, Phys. Rev. A \textbf{97}%
, 063814 (2018).

\bibitem{Tang} Y. Tang and W. Wang, Europhys. Lett. \textbf{58} (2),
188--194 (2002).

\bibitem{Dai2} C. Q. Dai, G. Q. Zhou, and J. F. Zhang, Phys. Rev. E \textbf{%
85}, 016603 (2012).

\bibitem{Dai3} C. Q. Dai, Y. Y. Wang, and J. F. Zhang, Opt. Lett. \textbf{35}%
, 1437 (2010).

\bibitem{Liu} R. Yang, R. Hao, L. Li, Z. Li, G. Zhou, Opt. Commun. \textbf{%
242}, 285--293 (2004).

\bibitem{Agrawal} G. P. Agrawal, Nonlinear Fiber Optics, 4th ed. (Academic,
Boston, 2006).
\end{thebibliography}
\end{document}